\documentclass[twocolumn,british,prl,superscriptaddress,showpacs]{revtex4-1}
\usepackage[letterpaper]{geometry}

\usepackage{graphicx}
\usepackage{amsmath}
\usepackage{amssymb}
\usepackage{esint}
\usepackage{epstopdf}
\usepackage{subfigure}
\newcommand{\bra}[1]{\langle #1|}
\newcommand{\ket}[1]{|#1\rangle}
\newcommand{\braket}[2]{\langle #1|#2\rangle}
\renewcommand{\t}[1]{\textrm{#1}}

\makeatletter
\@ifundefined{textcolor}{}
{%
 \definecolor{BLACK}{gray}{0}
 \definecolor{WHITE}{gray}{1}
 \definecolor{RED}{rgb}{1,0,0}
 \definecolor{GREEN}{rgb}{0,1,0}
 \definecolor{BLUE}{rgb}{0,0,1}
 \definecolor{CYAN}{cmyk}{1,0,0,0}
 \definecolor{MAGENTA}{cmyk}{0,1,0,0}
 \definecolor{YELLOW}{cmyk}{0,0,1,0}
 }

\makeatother

\makeatother

\usepackage{babel}

\begin{document}

\title{Matrix product states for quantum metrology}

\author{Marcin Jarzyna, Rafa{\l} Demkowicz-Dobrza{\'n}ski}
\affiliation{Faculty of Physics, University of Warsaw, ul. Ho\.{z}a 69, PL-00-681 Warszawa, Poland}

\begin{abstract}
We demonstrate that the optimal states in lossy quantum interferometry may be efficiently simulated using low rank matrix product states.
We argue that this should be expected in all realistic quantum metrological protocols with uncorrelated noise and is related
to the elusive nature of the Heisenberg precision scaling in the asymptotic limit of large number of probes.
%This opens up new possibilities of analyzing metrological protocols in the regime of large number of probes.
%We discuss the efficient way of finding almost optimal states for quantum metrology in the presence of decoherence both with certain measurements and %in general, using more sophisticated tools.
%We explicitly show cases  in which asymptotic bounds derived previously are or are not saturable.
%Eventually we show that spin squeezed states together with some specific measurement can reach asymptotic limit in most cases.
%COS TAM JESZCZE DODAC.
\end{abstract}

\pacs{03.65.Ta, 06.20Dk, 02.70.-c, 42.50.St}

\maketitle

Over the recent years, advancements in quantum engineering have pushed
non-classical concepts such as entanglement and squeezing, previously regarded as largely academic
topics, close to practical applications. Quantum features of light and atoms
helped to improve the performance of measuring devices that operate in the regime where the precision
is limited by the fundamental laws of physics \cite{Giovanetti}.
One of the most spectacular examples of practical applications of quantum metrology
can be found in gravitational wave detectors \cite{LIGO}, where
   the original idea \cite{Caves} of employing squeezed states of light to improve the sensitivity of an interferometer
   has found its full scale realization \cite{GEO, GEO2011}. No less impressive are
    experiments with trapped entangled ions demonstrating spectroscopic resolution enhancement crucial for the operation of the atomic clocks \cite{clock, Schmidt, Ospelkaus}.

When standard sources of laser light are being used, any interferometric
experiment may be fully described by treating each photon individually and claiming that \emph{each photon interferes only with itself}.
Sensing a phase delay $\phi$ between the two arms of the interferometer via intensity measurements
may be regarded as many independent repetitions of  single photon interferometric experiments.
 $N$ independent experiments result in the data that allows the parameter $\phi$ to be estimated with error scaling as
 $1/\sqrt{N}$---the so called standard quantum limit or the shot noise limit.
 If, however, an experiment cannot be split into $N$ independent processes, as is e.g. the case with the $N$ probing photons being entangled,
the above reasoning is invalid and one can in principle achieve the $1/N$  estimation precision---the Heiseberg scaling \cite{Giovanetti2004, Berry,
Lee, Zwierz}--- with
the help of e.g. the N00N states \cite{Dowling}.
\begin{figure}[t]
\includegraphics[width=\columnwidth]{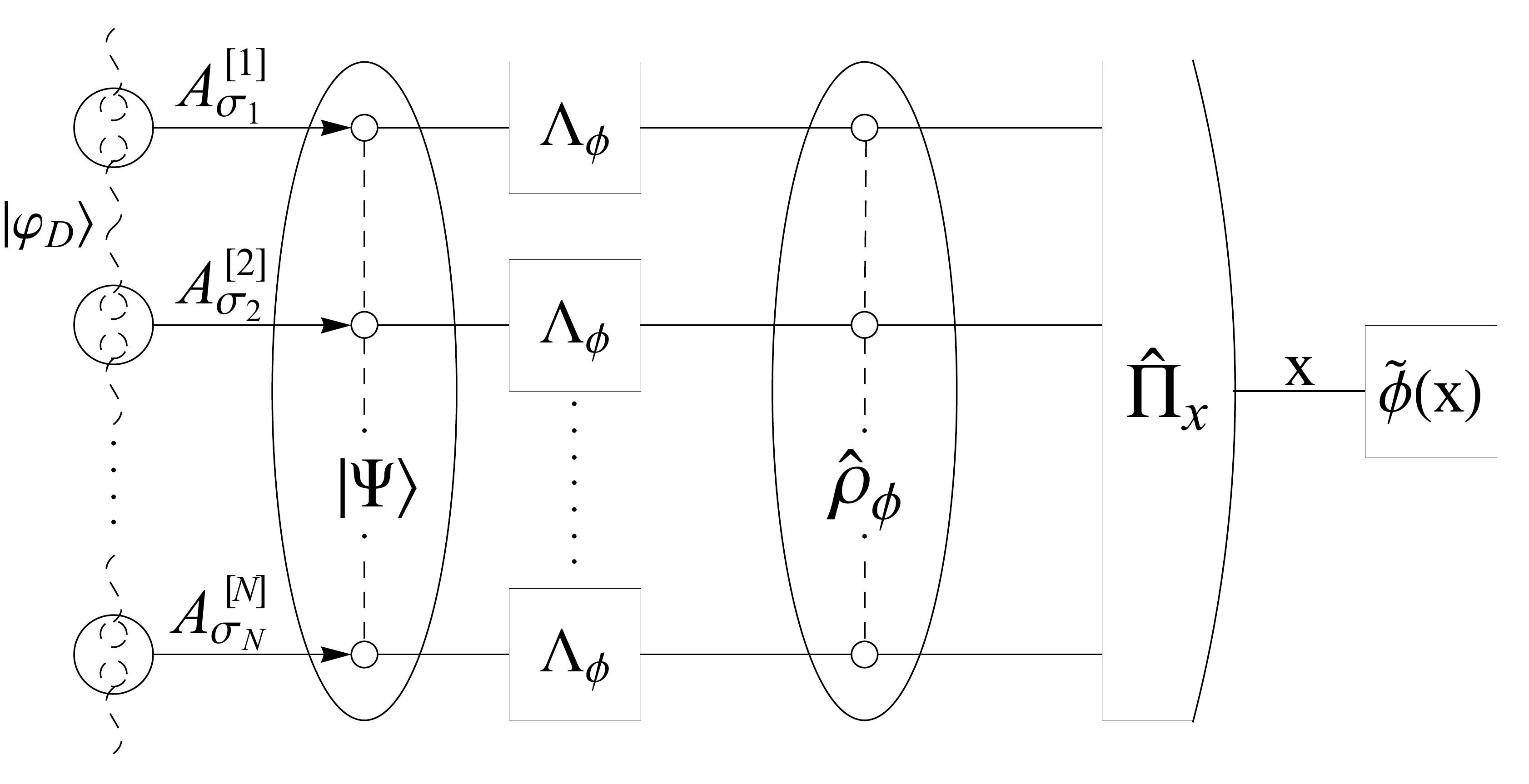}
\caption{Quantum metrology with matrix product states. $N$ parallel quantum channels act
on the input state $\ket{\Psi}$ inscribing parameter value $\phi$ as well as causing local decoherence on each of the probes.
Measurement $\hat{\Pi}_x$ on the output state $\hat{\rho}_{\phi}$ allows to make an estimate $\tilde{\phi}(x)$ based on the measurement result $x$.
The input state is a matrix product state obtained by the action of maps $A^{[k]}_{\sigma_{k}}$
on pairs of virtual $D$ dimensional systems prepared in such a way that the two adjacent systems corresponding to different neighboring particles are in a maximally entangled state $\ket{\varphi_{D}}$.}
\label{fig:bscr}
\end{figure}

Still, in all realistic experimental setups, decoherence typically makes the relevant quantum  features such as squeezing or entanglement die out very quickly \cite{Huelga, Shaji}. Recently, it has been rigorously shown for optical interferometry with loss \cite{Janek, Knysh}, as well as for more general decoherence models \cite{Escher,RafalJanek:Fujiwara}, that if decoherence acts independently on
each of the probes one can get at best $c/\sqrt{N}$ asymptotic scaling of error---precision that is better than classical one only by a constant factor $c$ which depends on the type of decoherence and its strength. One can therefore appreciate the Heisenberg-like decrease in uncertainty
only in the regime of small $N$, where the precise meaning of ``small'' depends on the decoherence strength \cite{Escher}, and
in typical cases is of the order of $10$ photons/atoms.

This indicates that in the limit of large number of probes, almost optimal performance can be achieved
by dividing the probes into independent groups where only the probes from a given group are entangled among each other. Clearly, the size of the group that is needed to approach the fundamental $c/\sqrt{N}$ bound up to a given
accuracy will depend on the strength of decoherence. Nevertheless, irrespectively of how small the decoherence strength is, for $N$ large enough
the size of the group will saturate at some point and therefore asymptotically the optimal state may be regarded as only locally correlated.

A natural class of states efficiently representing locally correlated states are the Matrix Product States (MPSs) \cite{MPS1,MPS3,MPS4,MPS5,MPS6},
which have proved to be highly successful in simulating low-energy states of complex spin systems. Until now no attempt has been made, however,
 to employ MPSs for quantum metrology purposes. Establishing this connection is the essence of the present paper.

Basic quantum metrology scheme is depicted in Fig.~\ref{fig:bscr}. $N$ probe input state $\ket{\Psi}$  travels through $N$ parallel noisy channels $\Lambda_\phi$ which action is parameterized by an unknown value $\phi$.
 A measurement $\hat{\Pi}_{x}$ is performed on the output density matrix $\hat{\rho}_{\phi}=\Lambda_\phi^{\otimes N}(\ket{\Psi}\bra{\Psi})$ yielding a result $x$ with probability $p(x|\phi)=\t{Tr}(\hat{\rho}_{\phi} \hat{\Pi}_x)$. The estimation procedure is completed by specifying
 an estimator function $\tilde{\phi}(x)$. Eventually we are left with the estimated value of the parameter, $\tilde{\phi}$, which in general will be
  different from $\phi$. We denote the average uncertainty of estimation by $\Delta \phi=\sqrt{\langle (\tilde{\phi} - \phi)^2 \rangle}$.
where the average is performed over different measurement results $x$.
   %and the deviation is quantified using a specific cost function $C_{\phi,\tilde{\phi}}$, e.g.the variance  $C_{\phi,\tilde{\phi}}=(\phi-\tilde{\phi})^2$.
   %From many types of possible decoherences, for quantum metrology typical are dephasing, depolarization, spontaneous emission, and losses of %particles form the setup. Here we will focus on the last one although similar results can be obtained in all other cases.
%To get optimal precision one has to optimize over the estimator, measurement, and also input state. Of course, in realistic experimental setups one can esily do only few specific kinds of measurement, like for example differnece of numbers of photons at the output of the interferometer. Nevertheless, still one gets different precisions for different input states. Finding optimal precision is thus finding an optimal state and this is hard task as in general it involves optimization over large number of parameters.
The main goal of theoretical quantum metrology is to find strategies that minimize $\Delta \phi$. For this purpose
one has to find the optimal estimator, measurement and input state. This in general is a difficult task.

To simplify the problem one may resort to the quantum Cramer-Rao inequality \cite{Braunstein, Helstrom, Nielsen, Paris}
\begin{equation}\label{eq:precF}
\Delta\phi\geq\frac{1}{\sqrt{k F(\hat{\rho}_\phi})}, \quad F(\hat{\rho}_\phi)= \t{Tr}( \hat{\rho}_\phi \hat{L}_\phi^2)
\end{equation}
that bounds the precision of any unbiased estimation strategy based on $k$ independent repetitions of an experiment.
$F(\hat{\rho}_\phi)$ is the Quantum Fisher Information (QFI) written
in terms of $\hat{L}_\phi$---the so called symmetric logarithmic derivative (SLD)---defined implicitly as:
 $2 \frac{d \hat{\rho}_\phi}{d \phi} = \hat{L}_\phi \hat{\rho}_\phi +\hat{\rho}_\phi \hat{L}_\phi$. For pure states
 the formula for QFI simplifies to $F(\ket{\Psi_\phi}) = 4(\braket{\dot{\Psi}_\phi}{\dot{\Psi}_\phi} -
 |\braket{\dot{\Psi}_\phi}{\Psi_\phi}|^2)$, where $\ket{\dot{\Psi}_\phi}= \frac{d \ket{\Psi_\phi}}{d \phi}$.
The bound is known to be saturable in the asymptotic limit
of $k \rightarrow \infty$  in the sense that there exist a measurement and an estimator that yields equality in \eqref{eq:precF}.
The main benefit of using QFI is that since it does not depend neither on the measurement nor on the estimator, the only remaining
optimization problem is the maximization of $F(\hat{\rho}_\phi)$ over input states.

Since the optimal states in the regime of large number of probes $N$ (not $k$)
 may be regarded as consisting of independent groups, the Cramer-Rao bound may be saturated even for $k=1$ provided $N$ is
 large enough \cite{Nielsen, Guta}. This makes the QFI an even more appealing quantity than in the decoherence-free case
 where some controversies arise on the practical use of the strategies based on the optimization of the QFI \cite{Anisimov, Giovanetti2012}.
Maximization of QFI over the most general input states for large $N$ may still be challenging, though, and even if successful might not provide
an insight into the structure of the optimal states. This is the place where MPSs come in useful.

A general MPS of $N$ qubits is defined as
\begin{equation}
\ket{\Psi}_{\textrm{MPS}}=\frac{1}{\sqrt{\mathcal{N}}}\sum_{\sigma_{1}\dots \sigma_{N}=0}^{1}\textrm{Tr}(A_{\sigma_{1}}^{[1]}\dots A_{\sigma_{N}}^{[N]})\ket{\sigma_{1}\dots \sigma_{N}},
\end{equation}
where $A_{\sigma_k}^{[k]}$ are square complex matrices of dimension $D\times D$, $D$ is called the bond dimension and $\mathcal{N}$ is the normalization factor.
%, $\mathcal{N}=\sum_{i_{1} \dots i_{N}=0}^{1}\textrm{Tr}[(A^*_{i_{1}}\otimes A_{i_{1}})\dots ({A}^*_{i_{N}}\otimes A_{i_{N}})]$.
In operational terms, a MPS is generated by assuming that each qubit is substituted by a pair of $D$ dimensional virtual systems. Adjacent systems corresponding to different neighboring particles are prepared in maximally entangled states $\ket{\varphi_{D}}=\frac{1}{\sqrt{D}}\sum_{\alpha=1}^{D}\ket{\alpha,\alpha}$ (Fig.~\ref{fig:bscr}) and
 maps $A_{\sigma_k}^{[k]}=\sum_{\alpha,\beta=1}^{D}A_{\sigma_k,\alpha,\beta}\ket{\sigma_k}\bra{\alpha, \beta}$ are applied to the pair of virtual systems
 corresponding to the $k$-th particle \cite{MPS5}.

Such a description of state is very  efficient provided the bond dimension $D$ increases slowly with $N$.
In a most favorable case when $D$ may be assumed to be bounded, $D < D_\t{max}$, the number of coefficients needed to specify an $N$ qubit state
in the asymptotic regime of large $N$ will scale as $N D_{\t{max}}^2$ (linear in $N$), as opposed to the standard $2^N$ scaling.
It should be noted, however, that
in many quantum metrological models, in particular the ones based on the QFI, the search for the optimal input probe states
may be restricted to symmetric (bosonic) states \cite{Huelga, Escher, Kolenderski, Rafal:OptStates}.
Even though the description of a symmetric $N$ qubit pure state is efficient and requires only $N+1$ parameters,
the use of MPSs may still offer a significant advantage as the symmetric MPS description
involves matrices $A$ which are identical for different particles: $A^{[k]}_\sigma = A_\sigma$ and
commute under the trace---$\t{Tr}(A_{\sigma_1} \dots A_{\sigma_N})$ does not depend on the order of matrices.
Provided $D$ is asymptotically bounded or grows slowly with $N$, one can still
benefit significantly from the use of MPS in the large $N$ regime.

In order to demonstrate the power of the MPS approach, we apply it to the most thoroughly analyzed and
relevant model in quantum metrology---the lossy interferometer.
We will not specify the nature of the physical systems (atoms, photons) but will rather refer to abstract two-level probes, with orthogonal sates $\ket{0}$, $\ket{1}$. % (e.g. a photon traveling in the lower/upper arm of the interferometer).
 %Consider $N$ distinguishable probes (e.g. photons prepared in different time bins), $\ket{\Psi} = \sum_{\sigma_1 \dots \sigma_N=0}^1 \alpha_{\sigma_1 %\dots \sigma_N} \ket{\sigma_1  \dots \sigma_N}$.
 The parameter to be estimated is the relative phase delay $\phi$ a probe experiences being in $\ket{1}$
 vs. $\ket{0}$ state. The decoherence
mechanism amounts to a loss of probes where each of the probes is lost independently of the others with probability $1-\eta$.
As such, this is an example of a general scheme depicted in Fig.~\ref{fig:bscr}.
Since the distinguishability of probes offers no advantage for phase estimation \cite{Rafal:OptStates} we move to the symmetric state description
where the general $N$ probe state
reads $\ket{\Psi} = \sum_{n=0}^{N}\alpha_n\ket{n,N-n}$, and $\ket{n,N-n}$ represents $n$ and $N-n$ probes in states $\ket{0}$ and $\ket{1}$ respectively.
%Formally speaking, switching from the distinguishable to the indistinguishable probes case is equivalent to restricting
%the full $N$ probe space to its fully symmetric subspace.
The output state $\hat{\rho}_{\phi}$ can be written explicitly as:
\begin{equation}
\label{eq:rho}
\hat{\rho}_\phi=\sum_{l_0=0}^{N}\sum_{l_1=0}^{N-l_0}p_{l_0 l_1}\ket{\Psi^{l_0 l_1}_\phi}\bra{\Psi^{l_0 l_1}_\phi}
\end{equation}
where
\begin{multline}
\ket{\Psi^{l_0 l_1}_\phi}=\frac{1}{\sqrt{p_{l_0 l_1}}}\sum_{n=l_0}^{N-l_1}\alpha_n e^{\mathrm{i} n \phi}\beta_{l_0 l_1}^{n}\ket{n-l_0,N-n-l_1}\\
\beta_{l_0 l_1}^{n}(\eta)= \sqrt{B_{l_0}^{n}B_{l_1}^{N-n}}, \ B_{l}^{n}={{n}\choose{l}}\eta^{n-l}(1-\eta)^{l}
\end{multline}
and $p_{l_0 l_1}$ is a normalization factor which can be interpreted as a probability to lose $l_0$, $l_1$ probes in states
 $\ket{0}$ and $\ket{1}$ respectively.

As the output state $\hat{\rho}_\phi$ is mixed, QFI is not explicitly given in terms of the input state parameters as it requires
involved calculation of the SLD or other equivalent quantities \cite{Escher, Toth, Fujiwara}.
In general, due to convexity, QFI for a mixed state is smaller than a weighted sum
of QFIs for pure states into which a mixed state may be decomposed. Nevertheless,
%since states $\ket{\Psi^{l_0 l_1}}$ with different total number of lost photons $l=l_0+l_1$ occupy orthogonal subspaces,
%the following  equality holds %$F(\hat{\rho}_\phi)=\sum_{l=0}^{N}F\left(\sum_{l_0=0}^{N-l}p_{l_0,l-l_0}\ket{\Psi^{l_0,l-l_0}_\phi}\bra{\Psi^{l_0,l-l_0}_\phi}\right)$.
it was shown in \cite{Dorner} that assuming additional knowledge of how many photons were lost
being in the state $\ket{0}$ and how many being in the state $\ket{1}$
does not improve the estimation precision appreciably. Hence, one can approximate the QFI with a weighted sum of QFIs all pure states entering the mixture $\hat{\rho}_\phi$:
\begin{equation}
\label{eq:fishapp}
F(\hat{\rho}_\phi) \lesssim \tilde{F}(\hat{\rho}_\phi) = \sum_{l_0=0}^N \sum_{l_1=0}^{N-l_0} p_{l_0,l_1} F(\ket{\Psi^{l_0,l_1}_\phi}).
\end{equation}
This approximation simplifies the calculations significantly since in our case %QFI for pure states has an explicit formula---in our case
$F(\ket{\Psi_\phi^{l_0l_1}}) = 4(\bra{\Psi_\phi^{l_0l_1}} \hat{n}^2  \ket{\Psi_\phi^{l_0 l_1}} - |\bra{\Psi_\phi^{l_0l_1}} \hat{n} \ket{\Psi_\phi^{l_0l_1}}|^2)$ with $\hat{n}$ being the the excitation number operator $\hat{n} \ket{n,N-n} = n \ket{n, N-n}$.
Direct optimization of formula \eqref{eq:fishapp} over the input state parameters $\alpha_n$
involves $N+1$ variables.
This approach was taken in \cite{Dorner, Rafal:OptStates}. Here we
consider the class of symmetric MPSs parameterized with two
diagonal (assuring commutativity) $D \times D$ matrices $A_{0}$, $A_{1}$.
%Even though for a given $D$ a larger class of symmetric states could be generated with more general matrices, we have checked that this does not %provide any computational benefit in our case.
These MPS are parameterized with $2D$ complex numbers, instead of $N+1$, and read explicitly:
\begin{equation}
\ket{\Psi}_{\textrm{MPS}}=\frac{1}{\sqrt{\mathcal{N}}}\sum_{n=0}^{N}\sqrt{{{N}\choose{n}}}\textrm{Tr}(A_{0}^n A_{1}^{N-n})\ket{n,N-n}.
\end{equation}
%Whenever possible one should compute
%the quantities of interest directly on the $A_i$ matrices.
%where QFI is expressed in terms of linear combination of variances of the excitation number operator $\hat{n}$
Thanks to the simple form of Eq~\eqref{eq:fishapp} it is possible to compute $\tilde{F}$ directly on $A_\sigma$ matrices and there is no need to go back to the less efficient standard description as would be the case with the formula \eqref{eq:precF}.

\begin{figure*}[t]
\includegraphics[width=1\textwidth]{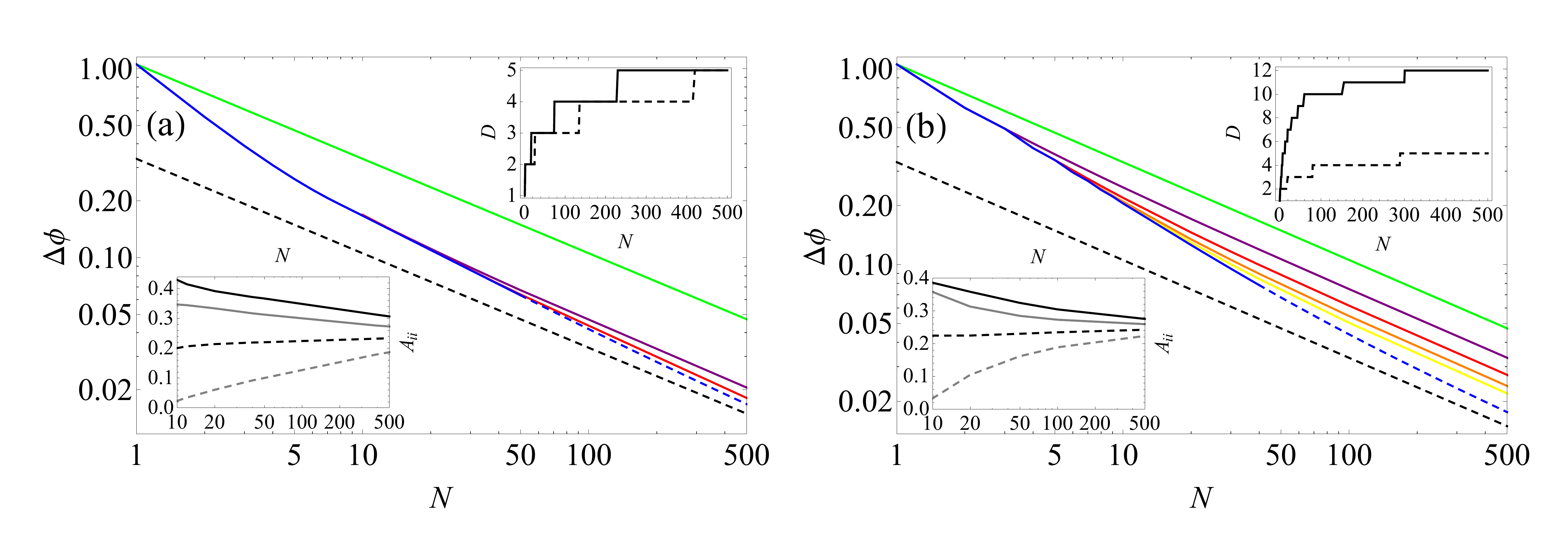}
\caption{(color online) Log-Log plots of the phase estimation precision with losses ($\eta=0.9$) as a function of
number of probes $N$ optimized over input states in (a) the QFI approach where $\Delta \phi = \frac{1}{\sqrt{F}}$ with $F$ given by Eq.~\eqref{eq:fishapp} and
(b) the Ramsey interferometry with $\Delta \phi$ given by Eq.~\eqref{eq:precLoss}.
The color curves correspond to matrix product states
of different bond dimensions $D$: $D=1$ (green, no correlations: $\Delta \phi=1/\sqrt{\eta N}$),
$D=2$ (purple), $D=3$ (red), $D=4$ (orange), $D=5$ (yellow).
The blue curves correspond to the optimal states
obtained with brute force optimization, with extrapolation for higher $N$ (dashed) obtained using scaling formulas of \cite{Knysh} (a)
and optimization over the spin-squeezed states \cite{Ma} (b). The black line
represents the asymptotic bound $\Delta \phi = \sqrt{\frac{1-\eta}{\eta N}}$ \cite{Janek,Knysh,Escher,RafalJanek:Fujiwara}.
The upper-right insets depict the bond dimension $D$ needed to reach the optimal precision with at most
$1\%$ discrepancy: $\eta=0.9$ (solid), $\eta=0.3$ (dashed).
Lower-left insets present normalized
absolute values of elements of the optimal diagonal MPS matrix $A_\sigma$ for bond dimension $D=4$ and $\eta=0.9$.}
\label{fig:combined}
\end{figure*}
%\begin{figure}[t]
%\includegraphics[width=1\columnwidth]{loss09ins2p.eps}
%\caption{(color online) Log-Log plot of the optimal phase estimation precision with losses ($\eta=0.9$) as a function of number of probes $N$ for matrix product states
%with different bond dimensions $D$: $D=1$ (green, no correlations: $\Delta \phi=1/\sqrt{\eta N}$), $D=2$ (purple), $D=3$ (red).
%The blue curve corresponds to the optimal states
%obtained with brute force optimization, with extrapolation for higher $N$ (dashed) obtained using scaling formulas of \cite{Knysh}. The black line
%represents the asymptotic bound $\Delta \phi = \sqrt{\frac{1-\eta}{\eta N}}$ \cite{Janek,Knysh,Escher,RafalJanek:Fujiwara}.
%The upper-right inset depicts the bond dimension $D$ needed to reach the optimal precision with at most $1\%$ discrepancy: $\eta=0.9$ (solid), $\eta=0.3$ (dashed). Lower-left inset presents normalized (normalization is to the sum of all values) absolute values of diagonal of optimal matrices $A$ for bond dimension $D=4$ and $\eta=0.9$.}
%\label{fig:lossMPS08}
%\end{figure}
 Fig.~\ref{fig:combined}a illustrates the precision obtained using MPS for the case of relatively small losses $\eta=0.9$.
  As one can see, the MPS approximation is excellent. In particular, the upper-right inset shows that already $D=5$ is sufficient
  to obtain less than $1\%$ discrepancy for $N \leq 500$.
We have confirmed this observation for different $\eta$ and observed %an intuitively understandable fact
that for higher losses (lower $\eta$) lower $D$ are required to obtain a given level of
approximation for a particular $N$---an effect that should be much more spectacular for larger $N$ reflecting the fact
that stronger decoherence diminishes the role of quantum correlations.

Moreover we have observed that optimal matrices $A_{0}$, $A_{1}$ have the same diagonal values which are ordered complementarily---the
 highest in $A_0$ is paired with lowest one in $A_1$ etc. The higher is $N$ the closer the diagonal values approach each other as can be seen from the lower-left inset on Fig.~(\ref{fig:combined}). This confirms the intuition that with increasing $N$, the optimal states
 are becoming less distinct from the product state---all diagonal values of $A_\sigma$ equal.

The peculiarity of phase estimation is that in the decoherence-free case optimal QFI is achieved for the N00N
state which, even though has non-local correlations, is an example of an MPS with $D=2$. This makes the MPS
capable of approximating the optimal states very well even for low loss and small $N$ $[N \lesssim 1/(1-\eta)]$---an ability that in general will not hold for other estimation problems.

%Until now we have focused our analysis only on the QFI and avoided discussion of any particular measurement scheme.
Taking now a more operational approach, not based on the QFI, one may consider a concrete measurement
scheme with a particular observable  $\hat{O}$ being measured.
Simple error-propagation formula for $\phi$ yields
$\Delta \phi = \Delta \hat{O} /|\frac{\t{d}\langle \hat{O}\rangle}{\t{d}\phi}|$.
In the Ramsey spectroscopy setup \cite{Wineland1}, or equivalently in the Mach-Zehnder interferometer with photon number difference measurement,
 one effectively measures a component of the total angular momentum operator $\hat{J}$ of $N$ spins 1/2---if a qubit $\ket{0}$, $\ket{1}$ is treated as a spin 1/2 particle \cite{Lee}. If the phase dependent rotation $\hat{U}=e^{i\phi\hat{J}_{z}}$ is being sensed by the measurement
 of the $\hat{J}_x$ observable, the explicit formula for estimation uncertainty at the optimal operation point $\phi=0$
 calculated for $\hat{\rho}_\phi$ from Eq.~\eqref{eq:rho} reads
\begin{equation}\label{eq:precLoss}
\Delta \phi=\sqrt{\frac{\Delta^2\hat{J}_x}{\langle\hat{J}_y\rangle^2}+\frac{1-\eta}{\eta}\frac{N}{4\langle\hat{J}_y\rangle^2}}.
\end{equation}
Search for the optimal state amounts to minimizing the above quantity. Since it  depends only on first and second moments of $\hat{J}$
it is simple to implement numerically using MPS. Results are presented in Fig.~\ref{fig:combined}b.

It is clear that MPS are capable to capture the essential feature  of the
optimal states---the squeezing of the $\hat{J}_x$---with relatively low bond dimensions $D$.
Moreover, the upper-right inset indicates that the required bond dimension $D$ is reduced much more significantly with increasing decoherence strength
than in the QFI approach.
The lower-left inset confirms again that the structure of the optimal states gets closer to the product state structure with increasing $N$.
We have also applied the MPS approach to Ramsey spectroscopy with other decoherence models including independent dephasing, depolarization and spontaneous emission and have obtained completely analogous results.
%\begin{figure}[t]
%\includegraphics[width=1\columnwidth]{ramsey_inset2p.eps}
%\caption{(color online)
%Log-log plot of precisione
%obtained in the particular phase estimation scheme based on Ramsey interferometric setup with loss ($\eta=0.9$)
%for input matrix product states with $D$: D=1 (green),
%D=2 (purple), D=3 (red), D=4., D=5, .
%The blue curve corresponds to optimization of Eq.~\eqref{eq:precLoss} over all input states which for large $N$ are very well approximated by
%the spin-squeezed states \cite{Ma} on which the extrapolation (dashed curve) is based.
%The black line represents the asymptotic bound $\Delta \phi = \sqrt{\frac{1-\eta}{\eta N}}$ \cite{Janek,Knysh,Escher,RafalJanek:Fujiwara}.}
%The same as in Fig.~\ref{fig:lossMPS08} but for a particular phase estimation scheme based on the Ramsey interferometry.
%Results are presented for matrix product states with bond dimensions $D=1$ (green), $D=2$ (purple), $D=3$ (red), $D=4$ (orange), D=5 (yellow).
%The blue curve corresponds to optimization of Eq.~\eqref{eq:precLoss} over all input states and for large $N$ is very well approximated by
%optimization over the spin-squeezed states \cite{Ma} on which the extrapolation (blue dashed curve) is based.}
%\label{fig:lossRamsey}
%\end{figure}

In summary, we have shown that MPS are very well suited for achieving the optimal performance in realistic quantum metrological setups and
 may reduce the numerical effort while searching for the optimal estimation strategies.
 Even though we have based our presentation on a single model of lossy phase estimation
we anticipate these conclusions to be valid in all metrological setups where decoherence makes the asymptotic Heisenberg scaling
unachievable---the intuitive argument being that no large scale strong correlations are needed to reach the optimal performance.
An intriguing open question remains: is it possible, as it is in many-body physics problems,
to obtain an exponential reduction in numerical complexity thanks to the use of MPS.
This is not possible when the optimal states
are known to be symmetric, as in the lossy phase estimation. % since then the Hilbert space dimension grows only linearly with $N$.
In problems, however, where distinguishability of probes is essential as e.g. in
Bayesian multiparameter estimation \cite{Chiribella, Bagan}, MPS might demonstrate their full potential when impact of decoherence is
taken into account.

%The only requirement for an efficient use of the MPS is to be able to express the estimation cost as an explicit function of $A_i$
%without the need of using the standard Hilbert space formalism.

We thank Janek Ko{\l}ody{\'n}ski and Konrad Banaszek for fruitful discussions and support.
This research was supported by the European Commission under the Integrating Project Q-ESSENCE,
Polish NCBiR under the ERA-NET CHIST-ERA project QUASAR
and the Foundation for Polish Science under the TEAM programme.

\end{document}